\documentclass[%
aps,
twocolumn,
superscriptaddress,
amsmath,amssymb,
prl,
floatfix,
]{revtex4-2}
\usepackage{url}
\usepackage{graphicx}
\usepackage{placeins}
\usepackage{dcolumn}
\usepackage{bm}
\usepackage[colorlinks, linkcolor=red, anchorcolor=green, citecolor=blue]{hyperref}

\usepackage{physics}
\usepackage{dsfont}
\usepackage{tikz,pgfplots}
\usepackage{enumerate}
\usepackage{subfigure}
\usepackage{bbold}
\usepackage{makecell}
\usepackage{array}
\usepackage{amsthm}

\newtheorem{Def}{Definition}
\newtheorem{thm}{\protect\theoremname}
\newtheorem{prop}[thm]{Proposition}

\newtheorem{lem}{Lemma}

\providecommand{\theoremname}{Theorem}

\usepackage{xcolor}

\begin{document}


\title{Spatial Incompatibility Witnesses for Quantum Temporal Correlations}

\author{Xiangjing Liu} 
\email{xiangjing.liu@ntu.edu.sg}
\affiliation{Nanyang Quantum Hub, School of Physical and Mathematical Sciences, Nanyang Technological University, 637371, Singapore}
\affiliation{Centre for Quantum Technologies, Nanyang Technological University, 637371, Singapore}

\author{Harshit Verma} 
 \affiliation{MajuLab, CNRS-UCA-SU-NUS-NTU International Joint Research Laboratory, 117543 Singapore}
 \affiliation{Centre for Quantum Technologies, National University of Singapore, 117543 Singapore}

\author{Yunlong Xiao}
\affiliation{Institute of High Performance Computing (IHPC), Agency for Science, Technology and Research (A*STAR), 1 Fusionopolis Way, \#16-16 Connexis, Singapore 138632, Republic of Singapore}

\author{Oscar Dahlsten} 
\email{oscar.dahlsten@cityu.edu.hk}
\affiliation{Department of physics, City University of Hong Kong, Tat Chee Avenue, Kowloon, Hong Kong SAR}
\affiliation{Institute of Nanoscience and Applications, Southern University of Science and Technology, Shenzhen 518055, China}

\author{Mile Gu} 
\email{mgu@quantumcomplexity.org}
\affiliation{Nanyang Quantum Hub, School of Physical and Mathematical Sciences, Nanyang Technological University, 637371, Singapore}
\affiliation{Centre for Quantum Technologies, Nanyang Technological University, 637371, Singapore}
 \affiliation{MajuLab, CNRS-UCA-SU-NUS-NTU International Joint Research Laboratory, 117543 Singapore}



\begin{abstract}  
We introduce a witness-based framework for certifying quantum temporal correlations via the pseudo-density matrix (PDM) formalism, which is a spatiotemporal generalization of the density matrix. We define spatial incompatibility (SI) as the minimum distance between a PDM and valid density matrices. For trace-norm distance, we show that this reduces to the PDM's negativity, enabling the construction of experimentally accessible SI witnesses. We derive a tight bound on SI for quantum channels and analyze the respective roles of state and channel coherence in witnessing SI. Our approach, unlike the Leggett–Garg (LG) framework, exploits incompatible measurements that generate coherence through state disturbance. We show that channels satisfying the LG inequality for incoherent states can still exhibit detectable SI, demonstrating that measurement disturbance enhances the certification of temporal correlations. 
\end{abstract}

\maketitle


\noindent{\emph{Introduction---}}Consider performing sequential measurements on a single quantum system at two distinct times. Each round produces correlated outcomes, recorded as a pair of random variables. These correlations capture the interplay between the system's dynamics, coherence, and the measurement backaction.  Quantum theory, in contrast to classical probability theory, allows for correlations between sequential measurements on a given system that are incompatible with correlations from simultaneous measurements on quantum subsystems. Certifying such correlations, i.e.\ being able to guarantee that the correlations were temporal, is therefore essential for understanding the quantumness of dynamical processes. Moreover, assessing the quantum nature of temporal correlations has found applications in quantum causal structure inference~\cite{liu2025quantum,PhysRevLett.130.240201,song2023causal}, partially inferring system dimension~\cite{vieira2024witnessing,spee2020genuine,assessing2009wolf}, and understanding the performance of quantum clocks~\cite{qclocks2022,ticking2021budroni}.

The analysis of temporal quantum correlations often takes inspiration from the analysis of spatial correlations. The Leggett-Garg (LG) inequality~\cite{leggett1985quantum}, a paradigmatic characterisation of temporal correlations, can be viewed as analogous to Bell inequality. The LG inequality can certify non-classical correlations under the restriction of a fixed repeated measurement~\cite{leggett1985quantum, emary2013leggett}. Analogues of the Tsirelson upper bound on Bell violation also exist~\cite{bounding2013budroni}. In further developing the analogy between spatial and temporal quantum correlations it is natural to consider the temporal analogue of entanglement witnesses~\cite{terhal2002detecting}, i.e., observables whose expectation values certify the presence of temporal quantum correlations. 

\begin{figure}
  \centering
  \includegraphics[scale=0.55]{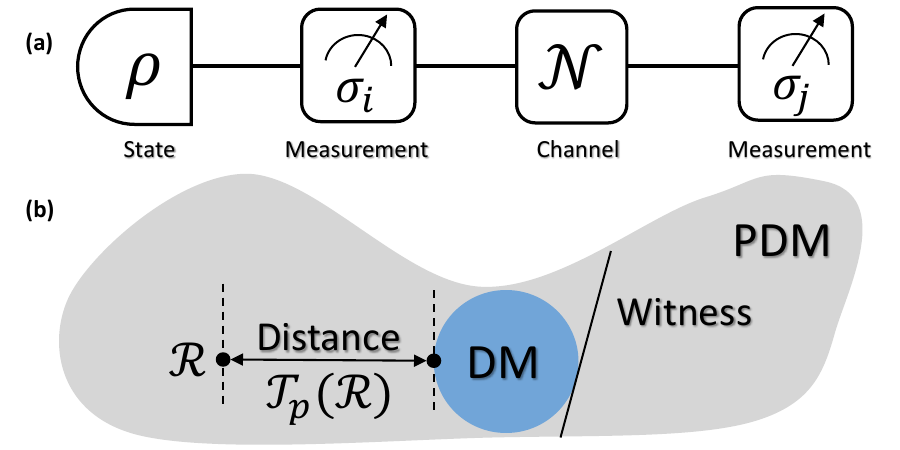}
  \caption{ {\bf{Setup.}} 
  (a) Measuring the two-time correlator $\langle \{\sigma_i, \sigma_j \} \rangle$.
  (b) Defining the measure of spatial incompatibility (SI) $\mathcal{T}_p(R)$  as the minimal distance between $R$ and the set of density matrices (DMs). The SI witness determines a hyperplane tangent to this set. The wider set of pseudo-density matrices (PDMs) is non-convex. 
 }
  \label{fig:setup}
\end{figure}

In this work, we establish a witness-based framework for certifying temporal quantum correlations. To such a purpose, we employ the pseudo-density matrix (PDM) formalism, which encodes spatiotemporal data~\cite{fitzsimons2015quantum}. We define a geometric measure of spatial incompatibility (SI) and construct experimentally accessible SI witnesses. We derive a tight bound on SI for CPTP dynamics and, within the resource theory of coherence together with the assistance of incompatible measurements, identify the roles of state and channel coherence in enabling detection. Finally, by comparing with the LG test, we reveal that measurement disturbance enables the certification of temporal correlations across a broader class of channels.

\smallskip
\noindent{\emph{Framework}---}We employ the PDM formalism~\cite{fitzsimons2015quantum}. The PDM extends the concept of the density matrix by associating a Hilbert space with each point in time and has found a wide range of applications in quantum information theory~\cite{marletto2019theoretical,marletto2021temporal,zhao2018geometry,quantum2024lie,lie2024unique,liu2024unification,parzygnat2023time,certifying2025liu}, such as quantum causal inference~\cite{inferring2024liu,liu2025quantum,song2023causal,song2025bipartite,liu2024experimental,liu2024quantumNMR}, estimating channel capacity~\cite{pisarczyk2019causal} and quantum broadcasting~\cite{virtual2024parzygnat,8vrg-tvsd,no2025xiaoL,no2025xiaoR}. In this work, we consider the PDM associated with two measurement events, which may be either space-like or time-like separated. Analogous to the tomographic construction of density matrices, the operational definition of the PDM in the multi-qubit system is given by
\begin{align}\label{eq: PDMdef}
    R_{12}=\frac{1}{4^n} \sum_{i_1,i_2} \langle \{\sigma_{i_1}, \sigma_{i_2}\} \rangle \sigma_{i_1} \otimes \sigma_{i_2},  
\end{align}
where $\sigma_i \in \{\mathbb{I}, \sigma_x, \sigma_y, \sigma_z\}^{\otimes n}$ denotes the multi-qubit Pauli observable and $\langle \{\sigma_{i_1}, \sigma_{i_2}\} \rangle$ represents the experimentally observed expectation value of the product of the corresponding measurement outcomes at two points. The construction of the two-time correlator is shown in Fig.~\ref{fig:setup}(a) and an example of a PDM is given later.  The PDM is Hermitian and has unit trace. But it may have negative eigenvalues. Moreover, the PDM can be generalized to arbitrary dimensions via a procedure analogous to density-matrix tomography~\cite{fullwood2024operator}.
 For simplicity, we present all the results for the multi-qubit system in the main body, with the generalization to $d$-dimensional systems provided in the Supplemental Material (SM).

\smallskip
\noindent{\emph{Definition of spatial incompatibility}---}We are interested in the features of quantum temporal correlations that are incompatible with quantum spatial correlations, i.e., with density matrices.
\begin{Def}[Spatial incompatibility] When the PDM $R$ induced by the statistics (via Eq.\eqref{eq: PDMdef}) is not a valid density matrix $\rho$ we say the statistics are spatially incompatible (SI). 
We define the ($p$-norm) degree of SI as 
\begin{align}
\label{eq:SIdegree}
    \mathcal{T}_p(R)= \min_{\rho \in \mathcal{D}}  || R-\rho||_p,
\end{align}
where $\mathcal{D}=\{\rho \succeq 0, \Tr \rho =1 \}$ is the set of density matrices and $||\cdot ||_p$ represents the Schatten-$p$ norm.
\end{Def}
By inspection, if $\mathcal{T}_p(R) >0$, then $R$ must exhibit temporal correlations; otherwise, it is compatible with quantum spatial correlations.
 
$\mathcal{T}_p(R)$, shown in Fig.~\ref{fig:setup}(b), has the following properties, with proofs provided in the SM:  I) Positivity: $\mathcal{T}_p(R) \geq 0 $; II) Convexity: $\mathcal{T}_p(\sum_i p_i R_i) \leq \sum_i p_i \mathcal{T}_p(R_i) $; III) Unitary invariance: $\mathcal{T}_p(URU^\dag )= \mathcal{T}_p(R)  $; IV)  Monotonicity for $p=1$, i.e., the trace norm:  $\mathcal{T}_1(\mathcal{N}(R)) \leq \mathcal{T}_1(R)$ holds for any completely positive and trace-preversing (CPTP) map $\mathcal{N}$; V) for $p=1$ the SI measure reduces to the causal monotone defined in Ref.~\cite{fitzsimons2015quantum}, which is the absolute value of the sum of the negative eigenvalues of $R$, up to a constant factor. The properties III and IV are mathematically appealing, with their physical interpretations yet to be explored. 
Property V provides an alternative interpretation of the causal monotone of~\cite{fitzsimons2015quantum}.
Given the physical interpretation and mathematical properties~\cite{ganardi2022hierarchy} (e.g., easy to compute) of $\mathcal{T}_{p=1}(R)$, we will adopt it (the $p=1$ case) as our measure of SI from now on.

The two established methods for detecting SI through the negativity of 
$R$ is PDM tomography and the randomized-measurement protocol for estimating the second-order moment of $R$~\cite{certifying2025liu}, both applicable in the absence of prior knowledge about the underlying process. In the following, we introduce a scheme for certifying SI that exploits knowledge about the process.

\smallskip
\noindent{\emph{Witnessing SI}---}Inspired by the entanglement witnesses~\cite{terhal2002detecting}, we define SI witnesses.

\begin{Def}[SI witness]
A Hermitian operator $W$ serves as a witness of SI if its expectation value is negative for some $R$s associated with some temporal processes but non-negative for all density matrices, that is, 
\begin{align}
\label{eq： witness}
    \Tr [ W \rho] \geq 0, \forall \rho, \   \text{and} \ \exists R, \,s.t. \,  \Tr [WR] = :\langle W\rangle_t <0.
\end{align}
\end{Def}

\noindent The first condition in the SI witness definition (that $\Tr [ W \rho] \geq 0$) is equivalent to positive semidefiniteness of $W$, such that $W$ is proportional to a state. This is an important difference between $W$ and entanglement witnesses, which by definition give negative expectation values for some entangled states, and thus have negative eigenvalues.
In conclusion, by measuring the SI witness observable $W$, we can determine whether a particular $R$ is temporal: if $\langle W\rangle_t <0$, we must conclude that it is not compatible with any quantum spatial correlations.

\smallskip
\noindent{{\em Constructing SI witnesses}---}Since we have chosen $\mathcal{T}_{p=1}(R)$, then, for $R$ to be spatially incompatible, it is equivalent to saying that $R$ has negative eigenvalues. Therefore, one can design the witnesses using the eigenvectors associated with those negative eigenvalues.  
Denote $\{\ket{E^-_n}\}_n$ and $\{\ket{E^+_m} \}_m$ as the sets of eigenvectors associated with the negative and non-negative eigenvalues, respectively. The straightforward way to construct a witness is to use any positive semidefinite operators in $\text{span} \{\ket{E_n^-}\}_n$. 
Generally, the witness can overlap with the space $\text{span} \{\ket{E_m^+}\}_m$ as long as it is constructed in a way such that $\langle W\rangle_t <0$. 

There is a subtlety in implementing the SI witness. The PDM $R$ here is a linear operator sitting in two Hilbert spaces at different times, i.e., $R\in \mathcal{L}(H_1 \otimes H_2) $, where the subscripts denote the time indices. The same applies to $W$. Similar to the construction of 
$R$, implementing the Hermitian operator $W\in \mathcal{L}(H_1 \otimes H_2)$ requires decomposing $W$ into Pauli operators that can be implemented locally in time, such as
   $ W= \sum a_{i_1j_2} \sigma_{i_1} \otimes \sigma_{j_2}.$
Such decomposition is always possible as the Pauli basis is complete. Therefore, the expectation $\langle W \rangle_t$ can be evaluated from the collection of the two-time correlators $\{ \Tr[ (\sigma_{i_1} \otimes \sigma_{j_2}) R] = \langle \{ \sigma_{i_1}, \sigma_{j_2} \} \rangle \}$. The negative eigenvalue of $R$ corresponds to a negative expectation value of an observable $W$. The above discussions can be summarized into the following proposition.
\begin{prop}[Existence of SI witness]
One can always design a positive semidefinite operator $W$, s.t., $\langle W \rangle_t <0$, whenever the PDM  $R$ is not positive semidefinite.
\end{prop}

Let us consider the following example to illustrate the construction and implementation of a witness. Consider the PDM associated with the state $\ketbra{0}{0}$ undergoing the identity channel $\mathcal{I}$, with measurements performed before and after the channel to obtain the two-time correlators. The corresponding PDM is $R=\frac{1}{4}\left( \mathbb{I}\otimes \mathbb{I}+ \sigma_x \otimes \sigma_x + \sigma_y \otimes \sigma_y+ \sigma_z \otimes \sigma_z +\mathbb{I} \otimes \sigma_z+ \sigma_z \otimes \mathbb{I}\right)$.
The eigenvector associated with the negative eigenvalue is $\ket{E^-} = \frac{1}{\sqrt{2}}(\ket{ 01} - \ket{10})$. Therefore, one can simply take $W=\ket{E^-}\bra{E^-}=\frac{1}{2} (\mathbb{I}\otimes \mathbb{I}- \sigma_{x_1} \otimes \sigma_{x_2} - \sigma_{y_1} \otimes \sigma_{y_2} - \sigma_{z_1} \otimes \sigma_{z_2})$, and collect the expectation values $\langle \{\sigma_{x_1}, \sigma_{x_2} \} \rangle, \langle \{\sigma_{y_1}, \sigma_{y_2}\} \rangle$ and $\langle \{\sigma_{z_1}, \sigma_{z_2} \} \rangle$ to evaluate $\langle W\rangle_t $.

\smallskip
\noindent{\emph{Bound on SI when the dynamics is CPTP}---}We now restrict our attention to those temporal processes described by CPTP maps.  Let $R$ be the PDM associated with a quantum state $\rho$,  which evolves from $t_1$ to $t_2$, with the intermediate evolution modeled by a quantum channel $\mathcal{N}$. 
 The construction of the two-time correlators is sensitive to the measurement scheme for multi-qubit Pauli observables. When choosing the projectors 
 \begin{align}\label{eq: measurement}
 \{P_+^i=(\mathbb{I}+\sigma_i)/2 , P_-^i= (\mathbb{I}-\sigma_i)/2\}
 \end{align}
 for $\sigma_i$ at each time point, i.e., projecting onto the $\pm1$ eigenspaces of $\sigma_i$,  the closed form of the PDM is given by~\cite{liu2025quantum}
\begin{align}\label{eq: closedPDM}
    R(\rho, \mathcal{N})=\frac{1}{2} \left\{ \rho \otimes \mathbb{I}, M_\mathcal{N} \right\},
\end{align}
where $\{ \cdot\}$ denotes the anticommutator and $M_{\mathcal{N}} = \sum_{i,j} \ket{i}\bra{j} \otimes \mathcal{N}(\ket{j}\bra{i})$ denotes the Jamio\l kowski matrix associated the channel $\mathcal{N}$~\cite{choi1975completely,jamiolkowski1972linear}. 

The maximal SI is of foundational interest in contexts such as generalized probabilistic theories~\cite{bounding2013budroni,popescu1994quantum}. We therefore derive the following proposition.

\begin{prop}[Maximal qubit SI degree] 
\label{prop: bound}
When the PDM $R$ has the closed form of Eq.~\eqref{eq: closedPDM}, qubit-to-qubit channels have the degree of SI bounded by
    \begin{align}
    \mathcal{T}_1(R(\rho, \mathcal{N})) \le      \mathcal{T}_1(R(\ketbra{0}{0}, \mathcal{I}))=1.
\end{align}
\end{prop}
\noindent The proof and the more general scenario are provided in SM. The proposition provides a tight upper bound on the quantumness of temporal correlations in qubit systems, clarifying the limit of certifiable temporal correlation within our setting.

\smallskip
\noindent{\emph{Role of channel in detecting SI with incoherent states}---}We are now in a position to discuss the role of quantum coherence in our approach to detecting quantum temporal correlations. Certifying temporal correlations can be regarded as a consequence of quantum mechanics, as it is classically impossible in general, captured by the phrase `correlation does not imply causation' in classical theory. Among the fundamental features of quantum mechanics is coherence, which we therefore adopt as the perspective for our analysis. Given the coarse-grained measurement scheme for the two-time correlators, the PDM $R(\rho, \mathcal{N})$ depends on both the input state $\rho$ and the channel $\mathcal{N}$. Therefore, we fix one of them to be `incoherent' and examine how coherence in the other influences the detection of SI. We first consider the case of incoherent initial states.

Let us begin by introducing the resource theories of quantum coherence. Both the state- and dynamics-based resource theories of quantum coherence are relevant to the study of quantum temporal correlations~\cite{smirne2018coherence}. In the state-based scenario, let $\{ \ket{i}\}$ be a fixed basis in the $2^n$-dimensional Hilbert space. The set of \emph{incoherent states} consists of all states that are diagonal in this basis, i.e., $\mathbb{D}:= \{\rho= \sum_i p_i \ket{i}\bra{i} \}$. To define incoherent operations, it is convenient to introduce the resource-destroying map~\cite{ziwen2017resource},  which in our case is the fully decohering operation $\Delta$. A channel $\mathcal{N}$ is \emph{creation-incoherent} (CI) if it cannot generate coherence from incoherent states, i.e., it satisfies
$\mathcal{N} \circ \Delta = \Delta \circ \mathcal{N} \circ \Delta$, where $\circ$ denotes the map composition. 
 The set of CI operations also constitutes the maximal set of incoherent operations.
In the dynamics-based scenario, if the aim is to detect coherence in the state, a channel $\mathcal{N}$ is called \emph{detection-incoherent} (DI) if the input coherence cannot impact the diagonal elements of the output state of the channel, i.e.,  it satisfies $ \Delta \circ \mathcal{N} = \Delta \circ \mathcal{N} \circ \Delta$~\cite{theurer2019quantifying}.

We are ready to investigate how the channel affects the detection of SI when the input state is incoherent. Our findings are summarized in the following Lemma.

\begin{lem}[Spatially compatible pairings of channels and incoherent input states]
\label{lem:IncohStat}
    Let $\rho = \sum_{i} p_i \ket{i}\bra{i} $ be a $n$-qubit incoherent state and $\Phi$ be a quantum channel acting on $\rho$. Consider the statistics of measurements as per Eq.~\eqref{eq: measurement}. Then, the two statements are equivalent: 
    1. The statistics are spatially compatible, i.e.,
    $R (\rho, \Phi)\succeq 0$ ;
    2. \, $\forall i, j=0,1,...,2^n-1$, the sub-matrix $R_{ij}$ is in the support of sub-matrix $R_{ii}$ and  $ R_{jj}-R_{ji}R_{ii}^\ddag R_{ij} \succeq 0$, where $R_{ij}= \frac{p_i+p_j}{2}\Phi(\ketbra{j}{i})$. 
\end{lem}
The proof is provided in the SM. The following proposition, which naturally follows from Lemma~\ref{lem:IncohStat} (see proof in the SM), identifies channels that are spatially compatible for any incoherent input.
\smallskip
\begin{prop}[OI channels and any fixed incoherent state spatially compatible]
\label{thm:OE}
A channel $\Phi$ is jointly spatially compatible (not SI) with {\em any} incoherent initial state and measurements as per Eq.~\eqref{eq: measurement} if and only if the channel $\Phi$ is off-diagonal independent (OI), i.e., $\Phi=\Phi \circ \Delta$.
\end{prop} 

\noindent  Channels that are spatially compatible with any incoherent state are a strict subset of incoherence operators in the dynamical resource theory of coherence.   Indeed, the OI channels characterized by Proposition~\ref{thm:OE} are DI operations as  
$\Phi=\Phi\circ\Delta$ implies $\Delta\circ\Phi=\Delta\circ\Phi\circ\Delta$, while the identity channel is DI but not OI.
However, OI channels might generate coherence: the proposition imposes no constraint on $\Phi(\ket{i}\bra{i})$, 
so an incoherent state can be mapped to a state with coherence. Therefore, the ability of channels to detect or generate coherence may not be necessary for detecting temporal correlations when incompatible measurements are involved.

We make a remark concerning the identity channel. We demonstrate that the identity channel serves as a valuable resource for detecting temporal correlations, just as it does in other quantum information processing tasks~\cite{horodecki2003entanglement,nico2024can,yuan2021universal}. However, its role should be understood in conjunction with measurement effects, as the identity channel combined with projective measurements can, in fact, generate coherence.

\smallskip
\noindent{\emph{Role of state coherence in detecting SI under classical channels.}---}The measurements associated with the two-time correlators can create coherence from incoherent states. For example, consider the measurement of $\langle \{ \sigma_x, \sigma_x\}\rangle$ in the example where the incoherent state $\ket{0}\bra{0} \in \mathbb{D}$ evolves under the identity channel. It proceeds as follows: the observable $\sigma_x$ is performed at time $t_1$, the resulting state is sent through the identity channel, and $\sigma_x$ is performed on the output. When $\sigma_x$ is measured at $t_1$, the post-measurement state becomes either $\ket{+}$ or $\ket{-}$, thereby creating coherence before the state enters the channel.

To probe whether quantumness in the input state is sufficient for detecting SI, given that the measurements can generate coherence, we restrict the channels to remove other quantumness. Specifically, we demand that the channels cannot detect coherence (OI) (c.f. Proposition~\ref{thm:OE}) and that they are \emph{coherence-erasing} (CE): $\mathcal{C}=\Delta \circ \mathcal{C}$. (The significance of the OI criterion will be discussed more later in the context of the LG inequality). CE channels will remove coherence from measurements. It is direct to see that CE channels are CI. The identity channel $\mathcal{I}$ is CI but not CE. 
 
We next show how coherence in the input state can enable the detection of SI even when the channel is classical, in the sense of being both CE and OI.
\begin{prop}[SI for coherent input states and classical channels]\label{thm: IncohChan}
There exist coherent input states undergoing certain simultaneously coherence-erasing (CE) and off-diagonal-independent (OI) channels, for which the statistics of measurements as per Eq.~\eqref{eq: measurement} are spatially incompatible (SI).
\end{prop}
\noindent The construction of the coherent state based on channels that are simultaneously CE and OI is provided in the SM, where a simple example is also given with a qubit in $\ket{+}$ undergoing the channel $\Delta$. Based on Proposition~\ref{thm: IncohChan}, for some classical channels, coherent states can make SI detectable. This demonstrates that coherence in the state plays a crucial role in certifying temporal correlations.

\smallskip
\noindent{\emph{Comparing with the LG test.}---}Here, we compare our framework with the LG test in detecting quantum temporal correlations. 

\begin{Def}(The LG inequality~\cite{leggett1985quantum}) 
Consider a dichotomic observable \( Q(t_i) = \pm 1 \) measured at different times \( t_1, t_2, t_3 \) and two-time correlator \( C_{ij} := \langle \{Q(t_i), Q(t_j)\} \rangle \). The LG inequality is that $C_{12} + C_{23} - C_{13} \le 1.
$
\end{Def}
\noindent The LG inequality follows from {\em macroscopic realism}: a macroscopic system with two or more macroscopically distinct states available to it will at all times be in one or the other of these states; and {\em noninvasive measurability}: it is possible, in principle, to determine the state of the system with arbitrarily small perturbation on its subsequent dynamics~\cite{leggett1985quantum}. Quantum mechanics can violate the inequality, suggesting the breakdown of at least one of these assumptions.

We first show that the LG test can also be employed to certify temporal correlations.  We formalize this point in the following proposition.
\begin{prop}[Spatial statistics cannot violate LG]
\label{prop:LGCertifying}
Measurements on three sites on a quantum system in a valid state $\rho$ cannot violate the LG inequality.
\end{prop}
\noindent The proof is provided in SM. Temporal correlations can simulate quantum entanglement~\cite{toner2003communication}. Here, we show that the converse is not true, i.e., the violation of the LG inequality cannot be simulated by any quantum spatial correlations, thereby establishing its ability to certify genuinely temporal correlations.

Secondly, the properties of channels that lead to violations of the LG inequality have been identified within the resource theory of coherence~\cite{smirne2018coherence,theurer2019quantifying,milz2020when}. It is done from a complementary viewpoint. The central idea is that a probability distribution via projective measurements that cannot be simulated classically, i.e., one that violates the Kolmogorov consistency conditions~\cite{breuer2002theory}, indicates genuinely quantum temporal correlations~\cite{emary2013leggett}. The noninvasiveness requirement is replaced by an assumption related to Markovianity. Consequently, the initial state is chosen to commute with the measured observable, that is, to be an incoherent state in the measurement basis. Under these conditions, to detect temporal correlation, i.e., violating the LG inequality, the channels must not be non-coherence-generating-and-detecting (NCGD)~\cite{smirne2018coherence}. Two important subsets of such channels are the CI and DI operations. More details are provided in SM.

\begin{figure}
  \centering
  \includegraphics[scale=0.22]{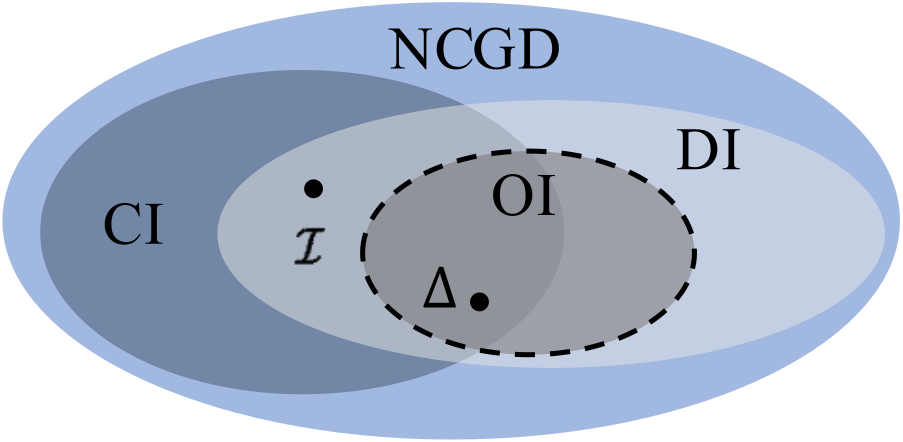}
  \caption{ {\bf{Hierarchy of Incoherent Operations.}}  
  The set NCGD contains both CI and DI operations as its subsets, with OI belonging to DI. 
  Operations inside OI cannot be detected by either SI or the LG inequality~\cite{smirne2018coherence}. 
  In contrast, operations outside OI but still within NCGD remain undetectable by the LG inequality yet can be detected by SI witness, demonstrating the broader capability of our approach in certifying temporal correlations.}

  \label{fig:adv}
\end{figure}

Finally, we summarize the comparison between the LG inequality and our witness-based approach in the following proposition.
\begin{prop} [Cases where LG is respected but SI witness exists]
\label{prop: advantage}
Consider the set of non-coherence-generating-and-detecting (NCGD) channels with subset off-diagonal independent ($OI$) channels. For channels in NCGD but not in OI: the LG inequality is respected for all incoherent initial states, while there exists at least one incoherent initial state such that an SI witness $W$ exists.
\end{prop}
 \noindent The relation between relevant operations is depicted in Fig.~\ref{fig:adv}. The key insight underlying Proposition~\ref{prop: advantage} is that certifying temporal correlations requires quantum resources. In contrast to the LG test, our framework introduces an additional source of quantumness: measurement disturbance. Measurements perturb the system’s state, thereby generating nonclassical resources and enhancing the ability to detect quantum temporal correlations.

\smallskip
\noindent{\emph{Summary and outlook}---}We proposed a witness-based framework for detecting quantum temporal correlations using incompatible projective measurements. We constructed experimentally accessible SI witnesses, temporal analogues of entanglement witnesses. We analysed the
roles of state and channel coherence in enabling detection showing, for example, that there can be SI for coherent initial states undergoing classical channels. For incoherent states, our approach, when compared with the LG test, reveals that state-disturbing measurements can enhance the certification of quantum temporal correlations.

Our work opens numerous avenues for future research. Firstly, it would be interesting to extend our results to a device-independent setting~\cite{vsupic2020self,chen2024semi}. Secondly, in the task of detecting quantum temporal correlations, a resource-theoretic framework that unifies coherence and measurement incompatibility is needed~\cite{incompatible2023}.  Thirdly, one may wonder whether more general measurements could offer an even better way to certify temporal correlations. Finally, an important direction is to generalize our results to encompass more complex temporal processes, particularly non-Markovian dynamics that cannot be adequately described by CPTP maps, but instead require the more general frameworks of quantum combs~\cite{theoretical2009chiribella}, process tensors~\cite{pollock2018non}, or process matrices~\cite{oreshkov2012quantum}.

  \smallskip
\noindent{\emph{Acknowledgements}---}We thank Ray Ganardi, Xueyuan Hu, Minjeong Song, Zhenhuan Liu, Qian Chen, Graeme Berk and Samyak Pratyush Prasad for helpful discussions.  X.L. and M.G. is supported by the National Research Foundation through the NRF Investigatorship on Quantum-Enhanced Agents (Grant No. NRF-NRFI09-0010) and the National Quantum Office, hosted in A*STAR, under its Centre for Quantum Technologies Funding Initiative (S24Q2d0009), the Singapore Ministry of Education Tier 1 Grant RT4/23 and RG77/22 (S), the FQXi R-710-000-146-720 Grant ``Are quantum agents more energetically efficient at making predictions?'' from the Foundational Questions Institute, Fetzer Franklin Fund (a donor-advised fund of Silicon Valley Community Foundation). H.V. is supported by the National Research Foundation, Singapore through the National Quantum Office, hosted in A*STAR, under its Centre for Quantum Technologies Funding Initiative (S24Q2d0009) and the Plan France 2030 through the project BACQ as part of the MetriQs-France program (Grant ANR-22-QMET-0002).
Y.X. acknowledges support from A*STAR under its Career Development Fund (C243512002). 
O.D. acknowledges funding from the City University of Hong Kong (Project No. 9610623)

  \smallskip
\noindent{\emph{Data availability}---}All data generated or analyzed during this study are available within the manuscript.

\bibliography{ref}

\appendix
\newpage

\begin{widetext}
\newpage

\setcounter{page}{1}
\setcounter{footnote}{0}
\thispagestyle{empty}

\begin{center}
	\textbf{\large Supplemental Material for \\
    ``Spatial Incompatibility Witnesses for Quantum Temporal Correlations''
}\\
	 \vspace{2ex}

\end{center}

\section{Pseudo Density Matrix for Arbitrary $d-$Dimensional Quantum Systems}
  Based on Ref.~\cite{fitzsimons2015quantum}, the operational definition of the pseudo density matrix (PDM) in qubit systems is given by
\begin{align}
    R_{12}=\frac{1}{4^n} \sum_{i_1,i_2} \langle \{\sigma_{i_1}, \sigma_{i_2}\} \rangle \sigma_{i_1} \otimes \sigma_{i_2},  
\end{align}
where $\sigma_i \in \{I, \sigma_x, \sigma_y, \sigma_z\}^{\otimes n}$ denotes the multi-qubit Pauli observable and $\langle \{\sigma_{i_1}, \sigma_{i_2}\} \rangle$represents the experimentally observed expectation value of the product of the corresponding measurement outcomes at two points. The definition can be generalized to arbitrary finite-dimensional quantum systems by following a procedure similar to the state tomography of finite-dimensional quantum systems. In the following, we provide the closed-form expression for the PDM associated with a multiqubit system and a quantum channel and generalize it to arbitrary finite-dimensional systems.

Let us consider the PDM associated with a multiqubit system and a CPTP map. In this case, the construction of the two-time correlators is sensitive to the measurement scheme for multi-qubit Pauli observables.  When choosing the projectors~\cite{liu2025quantum}
 \begin{align}
     \{P_+^i=(\mathbb{I}+\sigma_i)/2 , P_-^i= (\mathbb{I}-\sigma_i)/2\}
\end{align}
for $\sigma_i$ at each time point, i.e., projecting onto the $\pm1$ eigenspaces of $\sigma_i$, the two time correlator is given by~\cite{liu2025quantum}
\begin{align}
    \langle \{ \sigma_{i_1}, \sigma_{j_2}\} \rangle = \Tr [R(\rho, \mathcal{N}) (  \sigma_{i_1} \otimes \sigma_{j_2})],
\end{align}
where 
\begin{align}\label{eq: closedPDM1}
    R(\rho, \mathcal{N})=\frac{1}{2} \left\{ \rho \otimes \mathbb{I}, M_\mathcal{N} \right\},
\end{align}
with $\{ \cdot\}$ denoting the anticommutator and $M_{\mathcal{N}} = \sum_{i,j} \ket{i}\bra{j} \otimes \mathcal{N}(\ket{j}\bra{i})$ being the Jamio\l kowski matrix associated the channel $\mathcal{N}$~\cite{choi1975completely,jamiolkowski1972linear}.

The idea of obtaining the closed form of the PDM for qubit systems has been extended to arbitrary $d$-dimensional quantum systems~\cite{fullwood2024operator}. This is achieved by isolating key features of the Pauli operators, such as the fact that their spectra are always $\pm 1$ or $\pm \lambda$, leading to the introduction of \emph{light-touch observables}:
a light-touch observable is a Hermitian Operator $A$ such that the spectrum of $A$ is either $\{\lambda\}$ or $\{\pm \lambda\}$ for some $\lambda \ge 0$.

Consider a process with a quantum system of $d$-dimension and a quantum channel $\mathcal{N}$. Again, by choosing the projector that projecting onto the eigenvector space of $\pm \lambda$,
then, the two-time correlator can be calculated via 
\begin{equation}
\langle \{ A,B \} \rangle = \mathrm{Tr}\big[ R(\rho, \mathcal{N}) (A \otimes B) \big]
\tag{12}
\end{equation}
for all light-touch observables $A$ at the initial time and for all observables $ B$ at the final time with the expression of $R(\rho, \mathcal{N})$ given in Eq.~\eqref{eq: closedPDM1}. Therefore, if the set of light-touch observables  $\{A\}$ is tomographically complete, one can construct the PDM with the closed form given by Eq.~\eqref{eq: closedPDM1}.

In conclusion, all the results presented in this work apply to arbitrary finite-dimensional quantum systems.

\section{Properties of the Measure of Spatial Incompatibility}

\begin{prop}
    The proposed geometric measure of spatial incompatibility,
    \begin{align}
    \mathcal{T}_p(R)= \min_\rho || R-\rho||_p,
\end{align}
has the following properties:
\begin{enumerate}
    \item Positivity: $\mathcal{T}_p(R) \geq 0 $. 
    \item Convexity: $\mathcal{T}_p(\sum_i p_i R_i) \leq \sum_i p_i \mathcal{T}_p(R_i) $.
    \item Unitary invariance: $\mathcal{T}_p(URU^\dag )= \mathcal{T}_p(R)  $. 
    \item Monotonicity for $p=1$, i.e., the trace norm:  $\mathcal{T}_1(\mathcal{N}(R)) \leq \mathcal{T}_1(R)$ holds for any CPTP map $\mathcal{N}$.
    
\end{enumerate}
\end{prop}

\begin{proof}

We begin by noting that $\mathcal{D}$ is convex and compact, and the Schatten norm is continuous, hence the minimum is attainable.

\smallskip
\noindent\textbf{(i) Positivity.}
For any $\rho\in\mathcal D$, $\|R-\rho\|_p\ge 0$, whence $\mathcal T_p(R)\ge 0$.
If $R\in\mathcal D$, choosing $\rho=R$ gives $\mathcal T_p(R)=0$. Conversely, if $\mathcal T_p(R)=0$, then some minimizer $\rho^\star\in\mathcal D$ satisfies $\|R-\rho^\star\|_p=0$, i.e.\ $R=\rho^\star\in\mathcal D$.

\smallskip
\noindent\textbf{(ii) Convexity.}
Let $R=\sum_i p_i R_i$, with $p_i\ge 0$, $\sum_i p_i=1$. For each $i$, pick a minimizer
$\rho_i^\star\in\mathcal D$ with $\mathcal T_p(R_i)=\tfrac12\|R_i-\rho_i^\star\|_p$.
By convexity of $\mathcal D$, $\bar\rho:=\sum_i p_i\rho_i^\star\in\mathcal D$.
Then, using the triangle inequality,
\begin{align}
\mathcal{T}_p(R) &\le \big\|R-\bar\rho\big\|_p
 =  \Big\|\sum_i p_i(R_i-\rho_i^\star)\Big\|_p
 \nonumber\\
 &\le\; \sum_i p_i\|R_i-\rho_i^\star\|_p \nonumber\\
&= \sum_i p_i\,\mathcal T_p(R_i).
\end{align}

\smallskip
\noindent\textbf{(iii) Unitary invariance.}
Schatten norms are unitarily invariant: $\|UAV\|_p=\|A\|_p$ for unitaries $U,V$.
Moreover, $U\rho U^\dagger\in\mathcal D$ if $\rho\in\mathcal D$. Hence,
\begin{align}
\mathcal T_p(URU^\dagger)
&=\min_{\sigma\in\mathcal D}\|URU^\dagger-\sigma\|_p \nonumber\\
&=\min_{\rho\in\mathcal D}\|U(R-\rho)U^\dagger\|_p \nonumber\\
&=\mathcal T_p(R).
\end{align}

\smallskip
\noindent\textbf{(iv) Monotonicity for $p=1$.}
Let $\rho^\star$ attain $\mathcal T_1(R)=\|R-\rho^\star\|_1$.
Since $\mathcal N$ is CPTP, $\mathcal N(\rho^\star)\in\mathcal D$, and by
trace-norm contractivity under CPTP maps,
\begin{align}
\mathcal T_1(\mathcal N(R))
& =\min_{\sigma\in\mathcal D}\|\mathcal N(R)-\sigma\|_1 \nonumber\\
& \le \|\mathcal N(R)-\mathcal N(\rho^\star)\|_1 \nonumber\\
& \le \|R-\rho^\star\|_1 \nonumber\\
& =\mathcal T_1(R).
\end{align}
\end{proof}

We are now in a position to prove that for a special case of the $p$-norm, the spatial incompatibility reduces to the negativity of $R$.

\begin{prop}
    Let $p=1$, the measure of spatial incompatibility is given by the negativity of $R$, i.e.,
    \begin{align}
        \mathcal{T}_{p=1}(R):=\min_{\rho\in\mathcal D}\|R-\rho\|_1= 2\sum_{\lambda_i <0} | \lambda_i|, 
    \end{align}
    where $\lambda_i$ denotes the eigenvalue of $R$.
\end{prop}

\begin{proof}
Write the Jordan decomposition $R=R_{+}-R_{-}$ with $R_{\pm}\ge 0$ and $R_{+}R_{-}=0$.
Let $q:=\Tr R_{+}=\sum_{\lambda_i>0}\lambda_i$. Since $\Tr R=1$, we have
\begin{align}
    \Tr R_{-}=q-1=\sum_{\lambda_i<0}|\lambda_i|.
\end{align} 

\textbf{Upper bound by construction.}
Choose $\rho_0:=\frac1q R_{+}\in\mathcal D$. Using the orthogonality of the supports of
$R_+$ and $R_-$,
\begin{align}
\mathcal{T}_1(R) &\le \|R-\rho_0\|_1 \nonumber\\
& = \Big\|\Big(1-\frac1q\Big)R_{+}-R_{-}\Big\|_1 \nonumber \\
&=\!\left(\Big|1-\frac1q\Big|\,\|R_{+}\|_1+\|R_{-}\|_1\right) \nonumber\\
&=\big((q-1)+(q-1)\big) \nonumber\\
&=2(q-1).
\end{align}

\textbf{Lower bound.}
Let $\Pi_+$ be the projector onto the positive eigenspace of $R$. For any $\rho\in\mathcal D$,
set $X:=R-\rho$; then $X=X^\dagger$ and $\Tr X=0$. Using the variational
characterization,
\begin{align}
\|R-\rho\|_1
 &=2\max_{0\le P\le I}\Tr\!\big(P(R-\rho)\big)
\nonumber\\
&\ge\ 2\Tr\!\big(\Pi_+(R-\rho)\big) \nonumber\\
&= 2(\Tr R_+ - \Tr\!\big(\Pi_+\rho\big)).
\end{align}
Since $0\le \Pi_+\le I$ and $\Tr\rho=1$, we have
$$\Tr\!\big(\Pi_+ \rho\big)\le \Tr\rho=1.$$ Hence,
\begin{align}
\|R-\rho\|_1 \ \ge\ 2(\Tr R_+ - 1) \ =\ 2( q-1).
\end{align}
Taking the minimum over $\rho\in\mathcal D$ gives
\begin{align}
\min_{\rho\in\mathcal D}\|R-\rho\|_1 \ \ge\ 2(q-1).
\end{align}

Combining the two bounds yields \begin{align}
\mathcal T_1(R)=2(q-1)=2\sum_{\lambda_i<0}|\lambda_i|.
\end{align}
This completes the proof.
\end{proof}

This Proposition establishes the connection between spatial incompatibility and the causal strength monotone or quantum causal strength in the literature~\cite{fitzsimons2015quantum}, thereby providing an alternative explanation for the causal strength monotone.

\section{Bound on Spatial Incompatibility When the Dynamics is CPTP.}

 The maximal SI is of foundational interest in contexts such as generalized probabilistic theories~\cite{bounding2013budroni,popescu1994quantum}. Here we provide a bound for temporal correlation associated with processes that could be described by CPTP maps.\\ 

\smallskip
\noindent{\bf Proposition~\ref{prop: bound}.}
\emph{When the PDM $R$ is associated with a process that can be described by CPTP maps, it admits the closed form given in Eq.~\eqref{eq: closedPDM1}. For such a process corresponding to a CPTP map that acts on a 
$d$-dimensional input system and produces a $d$-dimensional output system, the SI is bounded by}
    \begin{align}
    \mathcal{T}_p(R(\rho, \mathcal{N})) \le      \mathcal{T}_p(R(\ket{\psi}, \mathcal{I})),
\end{align}
\emph{where $\ket{\psi}$ is a $d$-dimensional pure state and $\mathcal{I}$ is the identity channel.}

\begin{proof}
Let the channel $\mathcal{N}$ has the set of Kraus operators $\{K_l\}$ with the complete relation $\sum_l K_l^\dag K_l = \mathbb{I}$, then one has following decompositions
    \begin{align}
        M_{\mathcal{N}} &= \sum_{i,j,l} \ket{i}\bra{j} \otimes K_l(\ket{j}\bra{i})K_l^\dag  := \sum_l M_l \nonumber\\
        \rho&= \sum_k p_k\ketbra{\psi_k}{\psi_k}, \, \text{with} \, \sum_k p_k=1.
    \end{align}
      Plugging the two decompositions into  the closed-form of PDM, we have
    \begin{align}
            R(\rho, \mathcal{N}) &= \frac{1}{2} \left\{ \rho \otimes \mathbb{I}, M_\mathcal{N} \right\} \nonumber\\
            &= \frac{1}{2} \left\{ \sum_k p_k \ket{\psi_k}\bra{\psi_k} \otimes \mathbb{I}, \sum_l M_l  \right\} \nonumber\\
            &= \sum_{k,l} p_k R_{kl}.
    \end{align}
where $R_{kl}=\frac{1}{2} \left\{   \ket{\psi_k}\bra{\psi_k} \otimes \mathbb{I},  M_l  \right\}  $. 
    According to property II, the convexity of SI,  
    \begin{align}
    \mathcal{T}_p(\sum_i p_i R_i) \leq \sum_i p_i \mathcal{T}_p(R_i),
    \end{align}
    the maximal value of $\mathcal{T}_{1}(R(\rho, \mathcal{N}))$  is achieved when the state is pure and the dynamics are unitary. When the dynamics is described by a unitary channel $\mathcal{U}$, then one has the  Jamio\l kowski matrix as
    \begin{align}
        M_{\mathcal{U}} &= \sum_{i,j} \ket{i}\bra{j} \otimes U(\ket{j}\bra{i})U^\dag \nonumber\\
        &=  (\mathbb{I} \otimes U) M_\mathcal{I} (\mathbb{I} \otimes U^\dag),
    \end{align}
    where $\mathcal{I}$ denotes the identity channel.
    Based on the closed-form of PDM, we have
    \begin{align}
            R(\rho, \mathcal{U}) &= \frac{1}{2} \left\{ \rho \otimes \mathbb{I}, M_\mathcal{U} \right\} \nonumber\\
            &= \frac{1}{2} \left\{ \rho \otimes \mathbb{I}, (\mathbb{I} \otimes U) M_\mathcal{I} (\mathbb{I} \otimes U^\dag)  \right\} \nonumber\\
            &= (\mathbb{I} \otimes U )R(\rho, \mathcal{I}) (\mathbb{I} \otimes U^\dag).
    \end{align}
    Combining the above with property III $\mathcal{T}_p(URU^\dag )= \mathcal{T}_p(R)  $, one can choose any pure state and any unitary channel to attain the maximal value. Therefore we have 
    \begin{align}
            \mathcal{T}_p(R(\rho, \mathcal{N})) \le      \mathcal{T}_p(R(\ket{\psi}, \mathcal{I})).
    \end{align}

\end{proof}
  For an initial qubit state and a qubit-to-qubit channel, we choose the initial state to be $\ket{0}$ and the dynamics to be the identity channel.
    The corresponding PDM is 
    \begin{align}
    R(\ketbra{0}{0}, \mathcal{I})&=\frac{1}{4}\left( \mathbb{I}\otimes \mathbb{I}+ \sigma_x \otimes \sigma_x + \sigma_y \otimes \sigma_y+ \sigma_z \otimes \sigma_z +\mathbb{I} \otimes \sigma_z+ \sigma_z \otimes \mathbb{I}\right) \nonumber\\
 & =  \begin{pmatrix}
       1 & 0 &0 &0 \\
        0 & 0& 1/2 & 0 \\
        0 & 1/2 & 0 & 0 \\
        0 & 0 & 0& 0
    \end{pmatrix}. 
\end{align}
The set of eigenvalues of $ R(\ketbra{0}{0}, \mathcal{I})$ is $\{1,0, \pm 1/2\}$. Therefore, the SI is then bound by
\begin{align}
    \mathcal{T}_1(R(\rho, \mathcal{N})) \le      \mathcal{T}_1(R(\ketbra{0}{0}, \mathcal{I}))=2\times |-\frac{1}{2}|=1.
\end{align}

Therefore, the degree of SI is bounded by 1 for qubit-to-qubit channels.

\section{ Proof of Lemma~\ref{lem:IncohStat} and Proposition~\ref{thm:OE}}
\begin{lem}[Schur Complement]
\label{lem: one}
    Given any Hermitian matrix, $M= \begin{pmatrix}
        A & B \\
        B^\dag & C
    \end{pmatrix}$, the following conditions are equivalent ($\ddag$ means pseudo-inverse):
    \begin{enumerate}
        \item $M\succeq 0$ (positive semidefinite).
        \item  $A\succeq 0$, $(I-AA^\ddag)B=0$ \, ($B$ is in the support of $A$), $C-B^\dag A^\ddag B \succeq 0.  $
        \item $C\succeq 0, (I- CC^\ddag )B^\dag =0, A- BC^\ddag B^\dag \succeq 0.$
    \end{enumerate}
\end{lem}

\noindent{\bf Lemma~\ref{lem:IncohStat}.} \emph{  Let $\rho = \sum_{i} p_i \ket{i}\bra{i} $ be a $d$-dimensional incoherent state and $\Phi$ be a quantum channel acting on $\rho$. Consider the statistics of measurements as per Eq.~\eqref{eq: measurement}. Then, the following statements are equivalent:}
    \begin{enumerate}
        \item  \emph{The statistics are spatially compatible, i.e.,
    $R (\rho, \Phi)\succeq 0$ ;}
    \item \emph{  \, $\forall i, j=0,1,...,d-1$, (I) $R_{ij}$ is in the support of $R_{ii}$ and (II) $ R_{jj}-R_{ji}R_{ii}^\ddag R_{ij} \succeq 0$, where $R_{ij}= \frac{p_i+p_j}{2}\Phi(\ketbra{j}{i})$.}
    \end{enumerate}

\begin{proof}
Let $\Phi$ be the channel that maps the $d\times d$ density matrices to $n \times n$ density matrices, i.e., 
\begin{align}
\Phi : \mathcal{D}(H_1 )\to \mathcal{D}(H_2),
\end{align}
where $\mathcal{D}(H_1)$ and $\mathcal{D}(H_2)$ denote the sets of density matrices on the Hilbert spaces $H_1$ and $H_2$ of dimensions $d$ and $n$, respectively. Let $\rho =\sum^{d-1}_{i=0} p_i \ketbra{i}{i}$ be a diagonal state in the basis $\{ \ket{i}\}$.

First, consider $d=2$. Then, one obtain
\begin{align}
R(\rho = \sum_i p_i \ket{i}\bra{i}, \Phi)&= \frac{1}{2} \left( (\rho \otimes \mathbb{I}) M_\Phi + M_\Phi(\rho \otimes \mathbb{I})  \right) \nonumber\\
&= \sum^{1}_{i,j=0} \frac{(p_j+p_i)}{2}   \ket{i}\bra{j} \otimes \Phi(\ket{j}\bra{i})   \nonumber\\
&=\begin{pmatrix}
    R_{00} & R_{01} \\
    R_{10} & R_{11} 
\end{pmatrix},
\end{align}
where  $M_\Phi = \sum_{i,j} \ket{i}\bra{j} \otimes \Phi(\ket{j}\bra{i})$ denotes the Jamio\l kowski matrix associated the channel $\Phi$ and  
\begin{align}
R_{ij} = \frac{p_i+p_j}{2} \Phi(\ket{j}\bra{i}) \in \mathbb{C}^{n \times n} .
\end{align}
Note that $R_{ij} = R^\dag_{ji}$ since $R$ is Hermitian. 
The matrices $R_{ii}$ are positive semidefinite since $\Phi$ is a completely positive map. It follows from Lemma~\ref{lem: one} that $R\succeq 0$ if and ony if $R_{01}$ is in the support of $R_{00}$ and 
\begin{align}
    R_{11}- R_{10}R_{00}^\ddag R_{01}  \succeq 0 .
\end{align}
Therefore, the proposition holds for $d=2$.

For arbitrary $d>2$,
the PDM $R \in \mathcal{L}(H_1 \otimes H_2)$ can be expressed in terms of  a block matrix as follows,
\begin{align}
R(\rho, \Phi) 
= \begin{pmatrix}
        R_{00} & R_{01} & R_{02} &... \\
        R_{10} & R_{11} & R_{12} & ...\\
        R_{20} & R_{21} & R_{22} & ...\\
        \vdots & \vdots & \vdots & \vdots
    \end{pmatrix},
\end{align}
Again, the blocks $R_{ii}, i=0,1,...,d-1$ are positive semidefinite.
We first prove that $R(\rho, \Phi) \succeq 0$ if and only if 
\begin{align}
    R^{(i,j)}= \begin{pmatrix}
        R_{ii} & R_{ij} \\
        R_{ji} & R_{jj}
    \end{pmatrix} \succeq 0, \, \text{for} \ \   0\leq i\neq j \leq d-1.
\end{align} 

$\Rightarrow.$ 
Let us define 
\begin{align}
    \ket{\phi} = \ket{i} \otimes \ket{v_i}+ \ket{j} \otimes \ket{v_j}, \, \ket{v} \in \mathbb{C}^n.
\end{align}
It follows from $R \succeq 0$ that 
$
    \forall \ket{\psi} \in \mathbb{C}^{dn},  \bra{\psi} R \ket{\psi} \geq 0.
$
Therefore, we must have
\begin{align}
    \bra{\phi} R \ket{\phi} = \bra{\phi} R^{(i,j)} \ket{\phi} \geq 0, \forall \ket{\phi}.
\end{align}
This means that $R^{(i,j) } \succeq 0.$

$\Leftarrow$. We prove the statement by induction on $d$, using the Schur complement.
We have proved that the statement holds for $d=2$. Assuming it holds for $d\geq 3$, we aim to prove that it also holds for $d+1$.

Denote $R' \in \mathbb{C}^{nd \times nd}$. 
One can write 
\begin{align}
    R= \begin{pmatrix}
        R_{00} & C \\
        C^\dag & R'
    \end{pmatrix} \in \mathbb{C}^{n(d+1) \times n(d+1)},
\end{align}
where $C= [R_{01}, R_{02}, ..., R_{0d}]$.
Then for $R \succeq 0$, we need to prove that 
\begin{align}
    R_{00} \succeq 0, S:= R' - C^\dag R_{00}^\ddag C\succeq 0, (S\in \mathbb{C}^{nd \times nd}).
\end{align}

Based on our assumption, if we can prove that every $2 \times 2$ principal submatrix of 
$S$ is positive semidefinite, then $S\succeq 0$. 
To see this , fix $i,j \in \{1,2,...,d\}$, the corresponding $2\times 2$ block of $S$ is
\begin{align}
    S_{ij}
   = \begin{pmatrix}
        R_{ii}-R_{i0}^\dag R_{00}^\ddag R_{0i} & R_{ij}-R_{io}^\dag R_{00}^\ddag R_{0j} \\
         R_{ji}-R_{j0}^\dag R^\ddag R_{0i} & R_{jj}-R_{j0}^\dag R_{00}^\ddag R_{0j} 
    \end{pmatrix}
\end{align}
Now consider the $3 \times 3$ principal submatrix of $R$ on indices $\{0,i,j\}$:
\begin{align}
\begin{pmatrix}
R_{00} & R_{0i} & R_{0j} \\
R_{i0} & R_{ii} & R_{ij} \\
R_{j0} & R_{ji} & R_{jj}
\end{pmatrix} \succeq 0.
\end{align}
This $3 \times 3$ principle submatrix is positive semidefinite, as it falls within the scope of our assumption.
Applying the Schur complement of this $3 \times 3$ matrix with respect to the $(0,0)$ block $R_{00}$ yields the corresponding $2 \times 2$ block:
\begin{align}
\begin{pmatrix}
R_{ii} - R_{i0} R_{00}^\dagger R_{0i} & R_{ij} - R_{i0} R_{00}^\dagger R_{0j} \\
R_{ji} - R_{j0} R_{00}^\dagger R_{0i} & R_{jj} - R_{j0} R_{00}^\dagger R_{0j}
\end{pmatrix} \succeq 0.
\end{align}
Thus, every $2 \times 2$ principal submatrix of $S$ is positive semidefinite. By the induction hypothesis, $S \succeq 0$.
Therefore, we conclude that $R \succeq 0$.

In summary, $R \succeq 0$ iff $R^{(i,j)}\succeq 0 $,i.e., $R_{ij}$ is in the support of $R_{ii}$ and $ R_{jj}-R_{ji}R_{ii}^\ddag R_{ij} \succeq 0.$
\end{proof}

Lemma~\ref{lem:IncohStat} further implies that, given an incoherent state, there is SI ($R\not\succeq 0 $) for some channels. Importantly, the identity channel is one such channel. Indeed, for $\Phi=\mathcal{I}$, $R_{00} = p_0\Phi (\ket{0}\bra{0}) = p_0\ketbra{0}{0} (p_0 \neq 0)$ and $R_{01} = \frac{p_0+p_1}{2} \Phi(\ketbra{1}{0}) = \frac{p_0+p_1}{2} \ketbra{1}{0}.$  Since $\mathrm{supp}(R_{00})=\mathrm{span}\{\ket{0}\}$ while $\mathrm{range}(R_{01})=\mathrm{span}\{\ket{1}\}$, it is clear that $R_{01}\notin \mathrm{supp}(R_{00})$, so condition (I) fails.

The following proposition naturally follows from Lemma~\ref{lem:IncohStat}.\\

\smallskip
\noindent{\bf Proposition~\ref{thm:OE}.} \emph{A channel $\Phi$ is jointly spatially compatible (not SI) with {\em any} incoherent initial state and measurements as per Eq.~\eqref{eq: measurement} if and only if the channel $\Phi$ is off-diagonal erasing, i.e., $\Phi=\Phi \circ \Delta$.}

\begin{proof}
By the Lemma~\ref{lem:IncohStat}, for diagonal $\rho$, the PDM $R(\rho,\Phi)$ can be written in $d\times d$ blocks with
$
R_{ij}=\frac{p_i+p_j}{2}\,\Phi\!\big(|i\rangle\!\langle j|\big),
$
and $R(\rho,\Phi)\succeq 0$ if and only if every $2\times2$ principal block
$\begin{pmatrix} 
R_{ii} & R_{ij}\\ R_{ji} & R_{jj}
\end{pmatrix}$ is positive semidefinite.

\smallskip
\noindent(\(\Rightarrow\)) Assume $R(\rho,\Phi)\succeq 0$ for all diagonal $\rho$.
Fix $i\neq j$ and choose a diagonal state with $p_i=0, p_j>0$ and others arbitrary. Then
\begin{align}
R_{ii}=p_i\,\Phi(|i\rangle\!\langle i|)=0, \, 
R_{ij}=\frac{p_j}{2}\,\Phi(|i\rangle\!\langle j|).
\end{align}
Positivity of the block $\begin{pmatrix} R_{ii} & R_{ij}\\ R_{ji} & R_{jj}\end{pmatrix}
=\begin{pmatrix} 
0 & R_{ij}\\ R_{ij}^\dagger & R_{jj}
\end{pmatrix}$ implies $R_{ij}=0.$ 
Hence $\Phi(|i\rangle\!\langle j|)=0$. Since $i\neq j$ was arbitrary, the claim holds for all off-diagonal pairs.

\smallskip
\noindent(\(\Leftarrow\)) Conversely, suppose $\Phi(|i\rangle\!\langle j|)=0$ for all $i\neq j$. Then for any diagonal $\rho$,
$R_{ij}=0$ whenever $i\neq j$, while $R_{ii}=p_i\,\Phi(|i\rangle\!\langle i|)\succeq 0$ because $\Phi$ is completely positive.
Thus $R(\rho,\Phi)$ is block-diagonal with positive semidefinite diagonal blocks, and therefore $R(\rho,\Phi)\succeq 0$.

Next, we prove that the condition $\Phi(|i\rangle\!\langle j|)=0$ for all $i\neq j$ is equivalent to the condition $\Phi = \Phi \circ \Delta$.

\smallskip
\noindent(\(\Rightarrow\))  If $\Phi(|i\rangle\!\langle j|)=0$ for all $i\neq j$, then for any density matrix $\rho= \sum_{ij} p_{ij} \ket{i}\bra{j} $, we have
\begin{align}
    \Phi (\rho) = \sum_{i} p_{ii} \Phi(\ket{i}\bra{i}).
\end{align}
Since $\sum_i p_{ii} \ket{i}\bra{i} = \Delta (\rho)$, therefore, $\Phi(\rho) = \Phi (\Delta (\rho)) $.

\smallskip 
\noindent(\(\Leftarrow\)) Assume that $\Phi = \Phi \circ \Delta$. The for any $i\neq j$, 
\begin{align}
    \Phi(\ketbra{i}{j}) = \Phi(\Delta(\ketbra{i}{j}))=0.
\end{align}

\smallskip
This completes our proof.
\end{proof}

The off-diagonal erasing channels characterized by the above proposition can generate coherence. As an explicit example, let $\Phi$ have the set of Kraus operators
\begin{align}
  \{  K_0 = \ket{+} \bra{0},  K_1 = \ket{-}\bra{1}
\}.
\end{align}
Then, for any density operator $\rho$,
\begin{align}
\Phi(\rho) = \bra{0} \rho \ket{0} \ket{+}\bra{+} +  \bra{1} \rho \ket{1} \ket{-}\bra{-}.
\end{align}
Hence $\Phi(\ket{i}\bra{j})=0$ for $i\neq j$ (so $\Phi$ is DI), yet for the diagonal input $\rho=\ket{0}\!\bra{0}$ we have $\Phi(\rho)=\ket{+}\!\bra{+}$, which has coherence in the computational basis; therefore $\Phi$ is not CI.

\section{Proof of Proposition~\ref{thm: IncohChan}}

\noindent{\bf Proposition~\ref{thm: IncohChan}.} \emph{There exist coherent input states undergoing certain simultaneously coherence-erasing (CE) and off-diagonal-independent (OI) channels, for which the statistics of measurements as per Eq.~\eqref{eq: measurement} are spatially incompatible (SI).}

\begin{proof}
According to the properties of the channel $\mathcal{C}$, i.e., $\mathcal C= \mathcal C\circ\Delta$ and $\mathcal{C}=\Delta \circ \mathcal C$, one have
\begin{align}
\mathcal{C}( {\ketbra{i}{j} })=0, i\neq j \, \, \text{and}  \,\, \mathcal C(\ket{i}\!\bra{i})=\sum_{k=0}^{d-1} a_{k i}\, \ket{k}\!\bra{k},
\end{align}
where $ a_{k i}\ge 0$ and $ \sum_{k} a_{k i}=1.$ The Choi matrix of $\mathcal{C}$ is given by
\begin{align}
M_{\mathcal{C}}^T &=\sum_{i,j=0}^{d-1} |i\rangle\langle j |\otimes \mathcal C(|i\rangle\langle j|)  =\sum_{i,k=0}^{d-1} a_{k i}\ \ket{i}\!\bra{i}\otimes \ket{k}\!\bra{k},
\end{align}
The Jamio\l kowski matrix can be directly calculated by $ M_{\mathcal{C}} = (M_{\mathcal{C}}^T)^T= M_{\mathcal{C}} $.

For any state $\rho$, define $R_i:=\frac{1}{2}(\rho\ket{i} \bra{i}+\ket{i}\bra{i}\rho)$.
Then
\begin{align}
R(\rho,\mathcal C)
&= \frac{1}{2}\big((\rho\otimes \mathbb{I})M_{\mathcal C}+M_{\mathcal C}(\rho\otimes \mathbb{I})\big) \nonumber\\
&=\sum_{i,k} a_{k i} R_i \otimes \ket{k} \bra{k} \nonumber\\
&=\;\sum_{k} B_k \otimes \ket{k}\!\bra{k},
\end{align}
where $B_k:=\sum_{i} a_{k i}\,R_i$. Thus $R(\rho,\mathcal C)\succeq 0$ iff $B_k\succeq 0$ for all $k$.

Next we proof that if there exist indices $i\neq j$ and $k$ such that $a_{k i}\neq a_{k j}$,
then there exists a coherent state $\rho$ for which $R(\rho,\mathcal C) \not\succeq\;0 .$
Conversely, if all columns are identical, i.e.,\ $a_{k i}=b_k$ for every $i$,
then $R(\rho,\mathcal C)\succeq 0$ for all density operators $\rho$.

{\bf{Construction of a coherent input.}} 
Since $B_k$ is Hermitian, then, by Sylvester’s criterion, $B_k \succeq 0 $ is equivalent to all principal minors of $B_k$ are nonnegative. Motivated by this observation, we now construct a coherent input state that yields a nonzero SI.

Assume there exist $i\neq j$ and $k$ such that $a_{k i}\neq a_{k j}$. Restrict our attention to the two-dimensional subspace $\mathrm{span}\{\ket{i},\ket{j}\}$. 
Let
\begin{align}
\ket{\psi}=\sqrt{p}\,\ket{i}+ \sqrt{1-p}\,\ket{j},
\qquad 0<p<1,
\end{align}
then the operators $R_i$ and $R_j$ are
\begin{align}
R_i=\begin{pmatrix}
\frac{p}{2} & \frac{\sqrt{p(1-p)}  }{2}\\[4pt]
\frac{\sqrt{p(1-p)}  }{2} & 0
\end{pmatrix},
\qquad
R_j=\begin{pmatrix}
0 & \frac{\sqrt{p(1-p)} }{2}\\[4pt]
\frac{\sqrt{p(1-p)} }{2} & \frac{1-p}{2}
\end{pmatrix}.
\end{align}
Therefore the $k$-th block $B_k=a_{k i}R_i+a_{k j}R_j$ in this subspace equals
\begin{align}
B_k^{\{i,j\}}
=\begin{pmatrix}
\frac{a_{k i}}{2} p & \frac{a_{k i}+a_{k j}}{2}\, \sqrt{p(1-p)} \\[6pt]
\frac{a_{k i}+a_{k j}}{2}\,\sqrt{p(1-p)}  & \frac{a_{k j}}{2} (1-p)
\end{pmatrix}.
\end{align}
Its determinant is
\begin{align}
\det B_k^{\{i,j\}}
=\frac{p(1-p)}{4}\left(a_{k i}a_{k j}-\frac{(a_{k i}+a_{k j})^2}{4}\right)
=-\frac{p(1-p)}{16}\,(a_{k i}-a_{k j})^2\;<\;0.
\end{align}
Since $0<p<1$ and $a_{k i}\neq a_{k j}$,  $B_k$ has a negative eigenvalue,
so $R(\rho,\mathcal C)$ is not positive semidefinite.

{\bf{Channels that cannot have nonzero SI for all states.}}
If $a_{k i}=b_k$ is independent of $i$ for all $k$, then
\[
B_k=\sum_i b_k\,R_i=b_k\sum_i R_i
=b_k\,\frac{\rho+\Delta(\rho)}{2}\ \succeq\ 0,
\]
since $\rho$ and $\Delta(\rho)$ are positive.
Thus $R(\rho,\mathcal C)\succeq 0$ for all $\rho$.

This completes our proof.
\end{proof}

Let us consider a simple example. Consider the qubit case with $\mathcal{C} = \Delta $. 
This channel clearly satisfies $\mathcal C=\Delta \circ \mathcal C=\mathcal C \circ \Delta$. 
Choose the coherent input state 
\begin{align}
\ket{\psi}=\tfrac{1}{\sqrt{2}}(\ket{0}+\ket{1}), \quad 
\rho=\ket{\psi}\!\bra{\psi}.
\end{align}
The correponding Jamio\l kowski matrix and PDM  are given by
\begin{align}
M_{\mathcal C}
&=\sum_{i,j=0}^1 \ket{i}\!\bra{j}\otimes\mathcal C(\ket{j}\!\bra{i})
=
\begin{pmatrix}
1 & 0 & 0 & 0\\
0 & 0 & 0 & 0\\
0 & 0 & 0 & 0\\
0 & 0 & 0 & 1
\end{pmatrix}. \\
R(\rho, \mathcal{C})
&= \frac{1}{2}\big((\rho\otimes \mathbb{I})M_{\mathcal C}+M_{\mathcal C}(\rho\otimes \mathbb{I})\big) 
=
\begin{pmatrix}
1/2 & 0 & 1/4 & 0\\
0 & 0 & 0 & 1/4\\
1/4 & 0 & 0 & 0\\
0 & 1/4 & 0 & 1/2
\end{pmatrix}. 
\end{align}
Clearly, $R$ has negative eigenvalues.

Next, we follow the analysis in the proof to illustrate this example. For $k=0$, the corresponding $2\times 2$ block of $R(\rho,\mathcal C)$ is
\begin{align}
B_0 = \frac{1}{2} R_0 = 
\begin{pmatrix}
\tfrac14 & \tfrac14 \\[2pt]
\tfrac14 & 0
\end{pmatrix},
\qquad 
\det B_0 = -\tfrac{1}{16} < 0.
\end{align}
Hence $B_0$ has a negative eigenvalue, and therefore 
$
R(\rho,\mathcal C) \not\succeq 0.$

\section{Leggett–Garg Violation and Its Role in Certifying Temporal Correlations}

\noindent{\bf{Leggett-Garg inequality.}} The Leggett--Garg inequality is a fundamental test that probes whether the two assumptions can describe a physical system:
\begin{enumerate}
    \item \emph{macroscopic realism}: a macroscopic system with two or more macroscopically distinct states available to it will at all times be in one or the other of these states;
    \item \emph{noninvasive measurability}: it is possible, in principle, to determine the state of the system with arbitrarily small perturbation on its subsequent dynamics,
\end{enumerate}  
These are assumptions that hold in classical physics but may fail in quantum mechanics. Violation of the Leggett-Garg inequality therefore implies the breakdown of at least one of these classical assumptions, revealing intrinsically quantum features in temporal correlations. 

Consider a dichotomic observable \( Q(t_i) = \pm 1 \) measured at different times \( t_1, t_2, t_3 \). One defines the two-time correlation functions \( C_{ij} = \langle Q(t_i) Q(t_j) \rangle \). Under the two assumptions, these correlations must satisfy the Leggett--Garg inequality
\begin{align}
K = C_{12} + C_{23} - C_{13} \le 1.
\end{align}
 However, quantum mechanics allows violations of this bound, demonstrating the incompatibility between quantum dynamics and classical macrorealism.

To implement the Leggett-Garg inequality experimentally, one prepares a quantum system, e.g., a qubit, in an initial state, allows it to evolve under a known unitary, and performs sequential measurements of \( Q(t_i) \) at different times. The correlations \( C_{ij} \) are then reconstructed from ensemble statistics over repeated runs. Crucially, since direct sequential measurements can disturb the system, practical implementations use weak measurements, ideal negative-result measurements, or ancilla-assisted protocols to approximate noninvasiveness. 

\smallskip
\noindent{\bf{Conditions for the Leggett-Garg violation.}} The limitations of violating the Leggett–Garg (LG) inequality have been analyzed from the perspective of the resource theory of coherence~\cite{smirne2018coherence,theurer2019quantifying,milz2020when}. It is done from a distinct viewpoint: rather than testing the assumptions of macroscopic realism and noninvasive measurability, it seeks an operationally meaningful and clear-cut definition of classicality. The central idea is that a probability distribution obtained from projective measurements that cannot be simulated classically, i.e., one that violates the Kolmogorov consistency conditions~\cite{breuer2002theory}, reveals genuinely quantum temporal correlations~\cite{emary2013leggett}. Conversely, any set of probability distributions that satisfies the Kolmogorov conditions does not violate the corresponding Leggett–Garg inequalities.

In this framework, the noninvasiveness assumption is replaced by one related to Markovianity~\cite{smirne2018coherence}. Consequently, the initial state is chosen to commute with the measured observable, i.e., 
\begin{align}
[\rho_1, Q[t_1]] =0,
\end{align}
meaning that the initial state $\rho_1$ is an \emph{incoherent state} in the eigenbasis of the observable 
$O$.Under these conditions, a necessary condition for the violation of the Leggett–Garg inequality has been identified: the inequality can be violated only if the channels involved are coherence-generating-and-detecting (CGD). Its definition is as follows.
The channels which can be described  by the suitable Lindblad dynamics ${\Lambda(t) = e^{\mathcal{L}t}}, {t \in \mathbb{R}^+}$ are said to be CGD whenever there exist $t, \tau \in \mathbb{R}^+$ such that
\begin{equation}
\Delta \circ \Lambda(t) \circ \Delta \circ \Lambda(\tau) \circ \Delta
\neq
\Delta \circ \Lambda(t+\tau) \circ \Delta.
\end{equation}
 Otherwise, the dynamics is referred to as \emph{non-coherence-generating-and-detecting} (NCGD).
It is direct to see that the sets of CI (channels that satisfying $\mathcal{N} \circ \Delta = \Delta \circ \mathcal{N} \circ \Delta$) and DI (channels that satisfying $\Delta \circ \mathcal{N}  = \Delta \circ \mathcal{N} \circ \Delta$) operations are subsets of NCGD.
Therefore, CI and DI cannot violate the Leggett-Garg inequality.

\smallskip
\noindent{\bf{Its role in certifying temporal correlations.}} We are interested in which quantum temporal correlation is incompatible with spatial means instead of the quantum aspects of the temporal correlation. The following proposition establishes that the LG inequality can be served as a tool to certifying temporal correlation. 

\smallskip
\noindent {\bf Proposition~\ref{prop:LGCertifying}.}\emph{
Let $Q_1,Q_2,Q_3$ be $\{\pm 1\}$-valued Hermitian observables acting on three distinct subsystems, and let $\rho$ be any tripartite density operator on $\mathcal{H}_1\otimes\mathcal{H}_2\otimes\mathcal{H}_3$. Define}
\begin{align}
C_{12}:=\Tr \big[\rho\, (Q_1 \otimes Q_2 \otimes I)\big],\quad
C_{23}:=\Tr \big[\rho\, (I \otimes Q_2 \otimes Q_3)\big],\quad
C_{13}:=\Tr \big[\rho\, (Q_1 \otimes I \otimes Q_3)\big].
\end{align}
\noindent \emph{Then the Leggett--Garg combination satisfies}
\begin{align}
K  :=  C_{12}+C_{23}-C_{13} \le 1.
\end{align}
\emph{Consequently, any temporal dataset with $K>1$ cannot be reproduced by any tripartite quantum state \(\rho\) with local measurements $Q_1,Q_2,Q_3$.}

\begin{proof}
Let
\begin{align}
B := Q_1 \otimes Q_2 \otimes I  + I \otimes Q_2 \otimes Q_3  -  Q_1 \otimes I \otimes Q_3 .
\end{align}
Because $Q_1,Q_2,Q_3$ act on different tensor factors, the three terms in \(B\) pairwise commute and are jointly diagonalizable in a product eigenbasis of $Q_1,Q_2,Q_3$.
On a joint eigenvector with eigenvalues $q_1,q_2,q_3\in\{\pm1\}$, the eigenvalue of $B$ is
\begin{align}
\lambda(q_1,q_2,q_3)= q_1 q_2 + q_2 q_3 - q_1 q_3.
\end{align}
Therefore, the maximal value of $\lambda$ is 1 when for instance $q_1=q_2=q_3=1$, and the minimal value is -3 when, for instance, $q_1=q_3=1, q_2=-1$.
Therefore, for any density matrix $\rho$,
\begin{align}
-3 \le K \;=\Tr(\rho B) \;\le\;  1,
\end{align}
which proves the claim.
\end{proof}

In conclusion, the violation of the Leggett-Garg inequality, i.e., $K>1$,  is incompatible with any quantum spatial correlations, i.e., by correlations encoded in a quantum state.

\end{widetext}

\end{document}